\documentclass[floatfix,reprint,amsmath,amssymb,aps,prl]{revtex4-2}
\usepackage{graphicx}
\usepackage{bm}
\usepackage[hidelinks]{hyperref}
\usepackage[nameinlink,capitalize]{cleveref}
\usepackage{physics}

\newcommand{\gm}{\gamma_{-}}
\newcommand{\gp}{\gamma_{+}}
\newcommand{\avgS}[2]{\expval{S^{#1}}_{#2}} 

\begin{document}

\title{Macroscopic quantum synchronization effects}
\author{Tobias Nadolny}
\affiliation{Department of Physics, University of Basel, Klingelbergstrasse 82, 4056 Basel, Switzerland}
 \author{Christoph Bruder}
\affiliation{Department of Physics, University of Basel, Klingelbergstrasse 82, 4056 Basel, Switzerland}

\date{October 12, 2023}

\begin{abstract}
We theoretically describe macroscopic quantum synchronization effects occurring in a network of all-to-all coupled quantum limit-cycle oscillators.
The coupling causes a transition to synchronization as indicated by the presence of global phase coherence.
We demonstrate that the microscopic quantum properties of the oscillators qualitatively shape the synchronization behavior in a macroscopically large network.
Specifically, they result in a blockade of collective synchronization that is not expected for classical oscillators.
Additionally, the macroscopic ensemble shows emergent behavior
not present at the level of two coupled quantum oscillators.
\end{abstract}
\maketitle

In the presence of a sufficiently strong coupling, limit-cycle oscillators adjust their frequencies and exhibit coherence even in the presence of noise and disorder in their natural frequencies.
This phenomenon is called synchronization; it appears, e.g., in physical, biological, engineered, and social systems~\cite{10.1063/5.0026335,Buck,bridge}
and has been extensively studied in classical nonlinear dynamics~\cite{acebronKuramotoModelSimple2005,pikovsky_rosenblum_kurths_2001,sync,Okuda_1991,Montbrio_2004}.

Recently, understanding synchronization of quantum oscillators has attracted a great deal of interest~%
\cite{
Shepelyansky2006,
Chia2020,Lifshits2021,
ludwig_marquardt2013,lee_QuantumSynchronizationQuantum_2013,walterQuantumSynchronizationDriven2014,
rouletSynchronizingSmallestPossible2018,parra-lopezSynchronizationTwolevelQuantum2020,
PhysRevLett.125.013601,PhysRevA.105.062206,zhang2023observing,koppenhoferQuantumSynchronizationIBM2020,
Holland2014,zhuSynchronizationInteractingQuantum2015,
lee_QuantumSynchronizationQuantum_2013,Fazio2013,PhysRevE.107.024204,Witthaut2017,lorenzoQuantumSynchronisationClustering2021,Davis-Tilley_2018,Waechtler2023,SongHeshan2017b,ishibashiOscillationCollapseCoupled2017,
lorchGenuineQuantumSignatures2016, lorchQuantumSynchronizationBlockade2017,Cooper2019,Solano2019a,Chia2023,
PhysRevE.107.024204,
zhuSynchronizationInteractingQuantum2015, Witthaut2017,
Fazio2013,lorenzoQuantumSynchronisationClustering2021,
Zambrini2012,Fazio2013,Ameri2015,Armour2015,Bastidas2015,Armour2016,Weiss_2016,SongHeshan2017a,SongHeshan2017b,Talitha2017,WuJin-Hui2017,amitaiQuantumEffectsAmplitude2018,Kwek_Squeezing2018,Solano2019b,Jaseem2020,Fazio2020,Zambrini2021,Lutz2022,Parvinder2022,Biswabibek2022,BucaJaksch2022,KatoNakao2023,PhysRevA.99.033818,Davidsen2023,rouletQuantumSynchronizationEntanglement2018,koppenhoferOptimalSynchronizationDeep2019,jaseemQuantumSynchronisationNanoscale2020,solanki2022symmetries,
delmonte2023quantum,
PhysRevLett.131.030401,
Valencia-Tortora_2023}.
Quantum limit-cycle oscillators~\cite{Chia2020,Lifshits2021} can be implemented using harmonic 
oscillators~\cite{ludwig_marquardt2013,lee_QuantumSynchronizationQuantum_2013,walterQuantumSynchronizationDriven2014} or few-level systems~\cite{rouletSynchronizingSmallestPossible2018,parra-lopezSynchronizationTwolevelQuantum2020} supplemented by gain and loss.
Experimental realizations span systems of cold atoms~\cite{PhysRevLett.125.013601}, nuclear spins~\cite{PhysRevA.105.062206}, trapped ions~\cite{zhang2023observing}, and a simulation on a quantum computer~\cite{koppenhoferQuantumSynchronizationIBM2020}.
Large networks of quantum oscillators have been discussed, particularly two-level systems~\cite{Holland2014,zhuSynchronizationInteractingQuantum2015}
and harmonic oscillators~\cite{lee_QuantumSynchronizationQuantum_2013,ludwig_marquardt2013,ishibashiOscillationCollapseCoupled2017,PhysRevE.107.024204,Davis-Tilley_2018,SongHeshan2017b,Waechtler2023,Witthaut2017,lorenzoQuantumSynchronisationClustering2021,Fazio2013}.
In most cases, their synchronization is similar to that of classical noisy oscillators.
Some quantum features in such systems are discussed in Refs.~\cite{zhuSynchronizationInteractingQuantum2015,Witthaut2017,lorenzoQuantumSynchronisationClustering2021,Fazio2013,PhysRevE.107.024204},
e.g., the presence of entanglement and quantum discord.
Quantum effects beyond the influence of quantum noise that lead to a synchronization behavior qualitatively different from classical expectations have been studied mostly on the level of one, two, or three coupled oscillators~\cite{lorchGenuineQuantumSignatures2016, lorchQuantumSynchronizationBlockade2017,Solano2019a,Cooper2019,Chia2023}.

So far, it has remained an open issue whether quantum effects in synchronization survive when increasing the number of oscillators.
Will the quantum nature of the oscillators be reflected in the macroscopic dynamics?
Or does the detailed microscopic description of each oscillator become irrelevant resulting in large-scale dynamics described by generic classical synchronization models?
A third possibility is the emergence of behavior not visible at the level of few coupled oscillators.

In this work, we show how in a macroscopic ensemble of interacting quantum oscillators,
the synchronization behavior is qualitatively shaped by their quantum nature.
Both their wave-like character leading to interference and the discreteness of their energy levels
result in quantum synchronization effects that remain visible in a large network.
We explain these effects based on a comprehensive understanding of the behavior of each oscillator.
In contrast, we identify aspects of the dynamics that are understood as typical
synchronization transitions independently of the microscopic quantum properties.
Finally, we discuss phase frustration in the network: if the coupling causes each oscillator to favor antialignment of its phase with respect to the other oscillators, collective synchronization is suppressed. 
This results in emergent blockades of synchronization only present in the many-body system.

\textit{Model.---}
\begin{figure}
\includegraphics[width=3.4in,height = 1in]{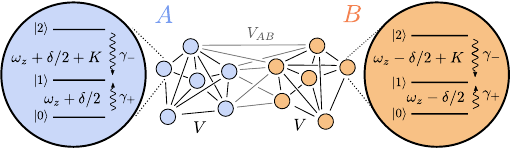}
\caption{\label{fig:system}
Two groups $A$ and $B$ of $N$ quantum oscillators each are all-to-all coupled through reactive interactions.
Each oscillator consists of three levels and is incoherently driven to state $\ket{1}$.
The asymmetry of the level structure is parameterized by $K$.
The oscillators in group $A$ are detuned from the ones in group $B$ by $\delta$.
}
\end{figure}
To address the issue of quantum synchronization effects in large networks of oscillators, we consider the minimal model schematically shown in \cref{fig:system}.
It comprises two groups of oscillators and thus resembles models of two ensembles of classical phase oscillators~\cite{Okuda_1991,Montbrio_2004}.
Here, however, each group consists of quantum oscillators with three states $\ket{0}$, $\ket{1}$, and $\ket{2}$ each.
In the frame rotating with the common frequency $\omega_z$ (see \cref{fig:system}), the time evolution is governed by the quantum master equation
$\dot \rho = -i[H_0 + H_\mathrm{int},\rho] +  \mathcal{L} \rho$, with
\begin{align}
    H_0 &=  \sum\nolimits_{i} \frac{\delta}{2} \left( S^z_{A, i} - S^z_{B, i} \right)
    +
    K\left( \dyad{2}{2}_{A,i}+\dyad{2}{2}_{B,i} \right) \nonumber
    \\
     H_\mathrm{int} &= 
     \frac{V}{N} \sum_{\sigma}\sum_{i < j}    S^+_{\sigma, i} S^-_{\sigma, j} +
     \frac{V_{AB}}{N}
     \sum_{i, j}
    S^+_{A, i} S^-_{B, j} 
    + \mathrm{h.c.} \nonumber
    \\
    \mathcal{L} &= \sum\nolimits_{\sigma, i}
    \left(
    \gp \mathcal{D} \left[\dyad{1}{0}_{\sigma,i} \right]
    +
    \gm \mathcal{D} \left[\dyad{1}{2}_{\sigma,i} \right]
    \right) \,.
    \label{eq:system}
\end{align}
In the sums, $i$ and $j$ take values from $1$ to $N$ and $\sigma \in \{A,B\}$ is the group label.
The spin--1 operators are defined as $S^z=\dyad{2}{2}-\dyad{0}{0}$, $S^+=\sqrt{2}(\dyad{2}{1}+\dyad{1}{0})$, and $S^- = (S^+)^\dag$.
The Lindblad dissipator is $\mathcal{D}[o]\rho = o\rho o^\dag -(o^\dag o \rho + \rho o^\dag o)/2$.

The bare Hamiltonian $H_0$ describes the coherent dynamics in the absence of any coupling. The two groups labeled $A$ and $B$ differ by the detuning $\delta$ between them.
The parameter $K$ sets the asymmetry in energy differences between levels $\ket{2}$ and $\ket{1}$, and the levels $\ket{1}$ and $\ket{0}$.
The interaction among the oscillators is described by $H_\mathrm{int}$.
All oscillators are reactively coupled to all others.
The coupling strength within each group is $V$, while the coupling strength between oscillators of different groups is $V_{AB}$.
Finally, each three-level oscillator is incoherently driven to the level $\ket{1}$, with strength $\gp$ ($\gm$) from level $\ket{0}$ ($\ket{2}$).
Due to these gain and loss processes, each three-level oscillator forms a limit cycle~\cite{rouletSynchronizingSmallestPossible2018}, whose population (measured by $S^z$) is stabilized, while the phase of the amplitude (measured by $S^+$) is free.

Because of the exponential growth of the Hilbert space size, solving the master equation becomes intractable for large $N$.
We employ a mean-field treatment, which for the case of an all-to-all coupling discussed here gives an exact solution for the macroscopic dynamics in the limit $N \rightarrow \infty$~\cite{RevModPhys.52.569}.
This approach corresponds to neglecting all correlations between operators, 
or, equivalently, using the product Ansatz,
$\rho = \bigotimes_{\sigma,i} \rho_{\sigma,i}$~\cite{leeEntanglementTongueQuantum2014}.
Since all oscillators within each group are identical, their time evolution can be described in terms of two three-level oscillators with density matrices $\rho_A$ and $\rho_B$ coupled to the mean amplitudes $\avgS{+}{\sigma} = \Tr[\rho_\sigma S^+] = 1/N \sum_i \expval{S_{\sigma,i}^+} $ of each group.
Consequently,
the dynamics of the two groups is described by the two coupled nonlinear master equations
\begin{align}
    \begin{split}
        \dot \rho_A &= 
        -i[H_A +
        V_{AB}(  S^+ \avgS{-}{B} + \mathrm{h.c.}) 
        ,
        \rho_A] +  \tilde{\mathcal{L}} \rho_A \, ,
    \\
        \dot \rho_B &= 
        -i[H_B +
        V_{AB}(  S^+ \avgS{-}{A} + \mathrm{h.c.})
        ,
        \rho_B] +  \tilde{\mathcal{L}} \rho_B \, ,
    \end{split}
    \label{eq:mf2}
\end{align}
where
$H_\sigma = \pm \frac{\delta}{2} S^z +
    K \dyad{2}{2} + V \left( S^+ \avgS{-}{\sigma} + \mathrm{h.c.} \right)$
and
$\tilde{\mathcal{L}} = 
    \gp \mathcal{D} \left[\dyad{1}{0} \right]
    +
    \gm \mathcal{D} \left[\dyad{1}{2} \right]
$.
The sign in front of $\delta/2$ is plus (minus) for group $A$ ($B$).
To obtain our results, we numerically time-integrate the nonlinear master equations~(\ref{eq:mf2}).
Additionally, we perform a stability analysis of the unsynchronized state $\rho_\sigma = \dyad{1}{1}$, which is a solution of $\dot \rho_\sigma = 0$. 
\begin{figure}[t]
 \includegraphics[width = 3.4in,height = 3in]{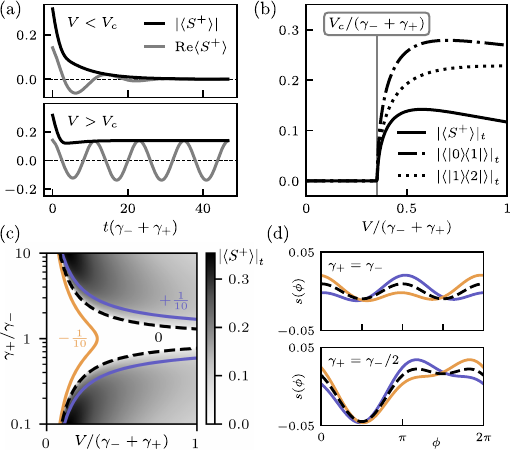}
\caption{\label{fig:1group}%
Synchronization of a single group.
(a)~Time evolution of the amplitude $\expval{S^+}$ below and above the critical coupling strength $V_c$.
(b)~Time-averaged long-time limit amplitude $\abs{\expval{S^+}}_t$ and coherences, showing a sharp transition at $V_c$.
Parameters in (a) and (b): $\gp=\gm/2$ and $K=0$.
(c)~Order parameter $\abs{\expval{S^+}}_t$ as a function of the coupling strength $V$ and $\gp/\gm$ as a gray-scale image for asymmetry parameter $K=0$.
The black dashed line displays the corresponding critical coupling strength obtained from a stability analysis.
The orange and blue lines show the critical coupling strengths for $K/(\gm+\gp) = -1/10$ and $ +1/10$, respectively.
In the region $\gp/\gm$ around 1, synchronization is suppressed for $K\ge 0$ due to the interference blockade; for negative values of $K$, the blockade vanishes and synchronization reappears, as indicated by the finite critical coupling strength for $K/(\gm+\gp) = -1/10$ (orange line).
(d)
Phase distributions $s(\phi)$ (the mean $1/2\pi$ is subtracted) for the same values of $K/(\gm+\gp)$ as in (c).
}
\end{figure}

\textit{Synchronization of a single group.---}
We begin to analyze the behavior of a single group by setting the inter-group coupling $V_{AB}$ to zero. For simplicity, the group subscript $\sigma \in \{A,B\}$ is omitted in this section.
To investigate the state of the group, we utilize the mean amplitude $\expval{S^+} = 1/N \sum_i \expval{S_i^+}$.
The phase $\phi_i$ of each oscillator is defined through $\langle S^+_i \rangle = \exp(i\phi_i) \abs{\langle S^+_i \rangle}$.
In the absence of any coupling, all oscillators exhibit random phases due to the intrinsic quantum noise.
Therefore, we expect the mean amplitude to vanish in the limit of infinitely many oscillators $N\rightarrow\infty$.
This conclusion follows from the mean-field analysis by noting that for small coupling strength, the group of oscillators converges to the steady state $\rho = \dyad{1}{1}$ exhibiting no phase preference since $\expval{S^+} = 0$.
The coupling among the oscillators, however, tends to align their phases.
As in the Kuramoto model describing classical phase oscillators~\cite{Kuramoto1975,acebronKuramotoModelSimple2005}, there is a critical coupling strength $V_c$ beyond which the group synchronizes.
The critical coupling usually depends on both the noise and the frequency disorder inherent in the system.
For a single group of identical oscillator, there is only intrinsic noise due to quantum fluctuations that increases with the rates $\gm$ and $\gp$ at which each oscillator couples to the environment.

\Cref{fig:1group}(a) displays the time evolution of the mean amplitude in both the unsynchronized and synchronized regimes. 
Below the critical coupling strength, the zero-amplitude state is stable.
For $V>V_c$, in the synchronized regime, the alignment of phases leads to a finite amplitude in the long-time limit with persistent oscillations of $\Re[\expval{S^+}]$.
The frequency of this oscillation will be further addressed when discussing the behavior of two coupled groups.
Other quantities not shown in \cref{fig:1group}(a) also change when entering the synchronized phase: the states $\ket{0}$ and $\ket{2}$ become populated, and the coherence $\expval{\dyad{0}{2}}$ exhibits oscillations at twice the frequency compared to those of $\expval{S^+}$.

To analyze the presence of synchronization among the oscillators, we use the time-average of the (in general time-dependent) absolute value of the amplitude $\abs{\expval{S^+}}_t$ in the steady state.
Figure~\ref{fig:1group}(b) depicts this order parameter as a function of the coupling strength, showing a sharp transition between the unsynchronized and synchronized states.
This resembles transitions to continuous time crystals~\cite{iemini_BoundaryTimeCrystals_2018,Tucker_2018} or to superradiance~\cite{Kirton_2019} that also result from the competition of coherent and incoherent dynamics in open quantum systems.

So far, we set $\gp/\gm=1/2$ and observed a typical synchronization transition.
We now present the order parameter as a function of both the coupling strength and the ratio $\gp/\gm$ in~\cref{fig:1group}(c).
Most notably, for equal gain and loss rates, the critical coupling diverges, i.e., the transition to synchronization disappears.
This is a macroscopic manifestation of the interference blockade.
Synchronization of a single three-level quantum limit-cycle oscillator subject to an external drive is suppressed when gain and loss rates are equal due to destructive interference~\cite{rouletSynchronizingSmallestPossible2018,koppenhoferOptimalSynchronizationDeep2019, solanki2022symmetries}.
Our result reveals that the wave-like character of the oscillators allowing for interference also shapes the dynamics in the macroscopic ensemble.

For a single asymmetric three-level oscillator, the interference blockade is lifted for \textit{any} non-zero value of the asymmetry parameter $K$~\cite{solanki2022symmetries}. 
In contrast, we find that the blockade in the macroscopic ensemble is lifted \textit{only} for $K<0$, but persists for $K\ge0$, see \cref{fig:1group}(c).
To understand this apparent discrepancy, we examine the microscopic quantum synchronization behavior.
\Cref{fig:1group}(d) shows the phase distribution $s(\phi)$ of a single oscillator coupled to an external drive in the steady state for various values of $K$.
The distribution $s(\phi)$ indicates the phase response of the driven oscillator~\cite{supp}.
In the case $\gamma_+ = \gamma_-$ and $K = 0$, the phase shifts $\phi = 0$ and $\phi = \pi$ between oscillator and drive are equally likely.
For the ensemble of oscillators, this  implies that any mean field present causes each oscillator to align itself both in and out of phase with the mean field.
Hence, the response of each oscillator will not amplify the coherence of the group,
which leads to the absence of synchronization, i.e., the macroscopic interference blockade.
For $K<0$, however, each oscillator preferably aligns its phase with the mean field leading to synchronization of the group. On the other hand, for $K>0$, each oscillator favors a phase shift of $\pi$ with respect to the mean field, resulting in \textit{phase frustration} that hinders synchronization.
Therefore, unlike the interference blockade of a single oscillator, the macroscopic interference blockade is only lifted for negative $K$.
For $\gp\neq\gm$, we similarly find that the phase distribution tends to peak closer to $\phi = 0$ ($\pi$) for negative (positive) values of $K$, which is reflected by the respective critical coupling strengths shown in \cref{fig:1group}(c) being smaller (larger).

In summary, the ensemble of quantum oscillators may synchronize above a critical coupling strength, as expected from
generic models of noisy classical oscillators.
Importantly though, the quantum nature of the oscillators remains influential on the macroscopic scale: destructive interference manifests itself as a blockade of global synchronization.
Moreover, phase frustration causes an emergent additional blockade only present in the large network.

\textit{Synchronization of two groups.---}
We now consider the full model where one half of the oscillators is detuned by $\delta$ from the other half.
We identify three different states in the long-time limit.
The first is the absence of any synchronization, indicated by both amplitudes $\avgS{+}{\sigma}$ vanishing.
Secondly, all oscillators of both groups can fully synchronize.
Thirdly, there is a state of partial synchronization where all oscillators within each group synchronize internally but not with the oscillators of the other group.

\begin{figure}
\includegraphics[width=3.4in,height=2.266666in]{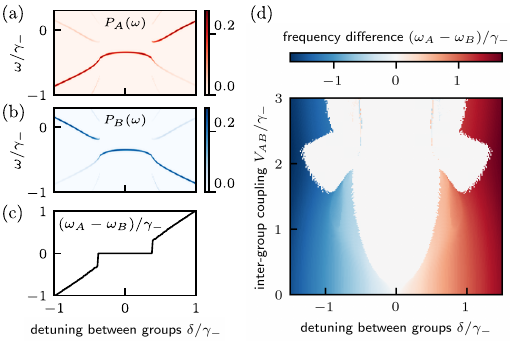}
 \caption{\label{fig:2groupsAnh0}
Synchronization of two groups.
(a) and (b)~%
Spectra $P_{\sigma}(\omega)$ for $\sigma=A,B$ obtained via Fourier transform of the time evolution of $\expval{S^+}_\sigma$ as a function of detuning $\delta$.
(c)~%
Difference of the two dominant frequencies, $\omega_A - \omega_B$.
(d)~%
Frequency difference between the two groups as a function of detuning $\delta$ and inter-group coupling $V_{AB}$.
Parameters: $K=0$, $V=2\gp=\gm$ (such that $V>V_c$). (a), (b) and (c): $V_{AB}=V/2$.
}
\end{figure}

To distinguish full and partial synchronization, we compare the oscillation frequencies of both groups.
For this purpose, we compute the discrete Fourier transform of the amplitudes in the long-time limit to obtain the spectra $P_{\sigma}(\omega)=|\mathrm{FT}\{\expval{S^{+}}_{\sigma}(t)\}|$ for each group.
\Cref{fig:2groupsAnh0}(a,b) displays the spectra as a function of the detuning $\delta$.
We set $V>V_c$ such that the oscillators are synchronized within each group.
For sufficiently small detuning compared to the inter-group coupling strength, we find a fully synchronized state as indicated by the identical spectra in this regime.
Since each spectrum is dominated by one frequency, we continue the analysis using the two frequencies at which the spectra peak, $\omega_{\sigma} = \mathrm{argmax}_\omega P_\sigma(\omega)$.
The frequency difference $\omega_{A} - \omega_{B}$ between the two groups is displayed in~\cref{fig:2groupsAnh0}(c).
At small $\delta$, the frequencies are equal and the two groups are synchronized, while for large detunings, $\omega_A$ and $\omega_B$ differ by $\delta$.
This corresponds to the dynamics described by the Adler equation~\cite{adler,pikovsky_rosenblum_kurths_2001} for classical phase oscillators.
To further demonstrate this correspondence, we show the frequency difference in \cref{fig:2groupsAnh0}(d) as a function of the inter-group coupling strength.
For $V_{AB} < V$,
both individually synchronized groups of oscillators can be regarded as two large oscillators which synchronize when their coupling is larger than their detuning.
In this regime, the microscopic details are irrelevant and the behavior 
matches that of generic synchronization models.
Specifically, we observe an Arnold tongue~\cite{pikovsky_rosenblum_kurths_2001}, i.e., the locking range grows with increasing coupling strength.
For $V_{AB} > V$, the inter-group coupling dominates, so that the analogy of two large coupled oscillators fails and the spectra show more than one relevant frequency component~\cite{supp}.
\begin{figure}
\includegraphics[width =3.4in,height = 2.7in]{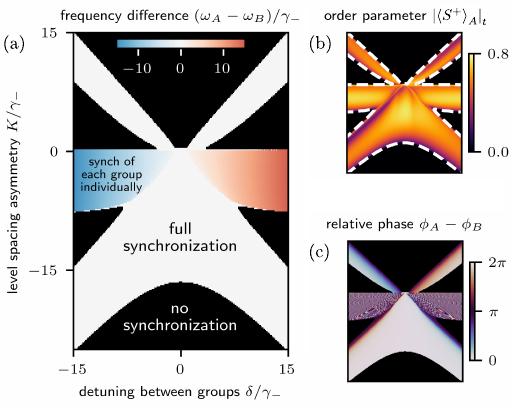}
\caption{\label{fig:2groupsVfix}
Macroscopic quantum synchronization blockade.
Panels (a), (b), and (c) show the frequency difference, the order parameter and the relative phase, respectively, as a function of detuning $\delta$ and asymmetry parameter $K$.
The phase $\phi_\sigma$ for $\sigma=A,B$ is the argument of $\expval{S^+}_\sigma$ in the long-time limit.
In (b) and (c), $\delta$ and $K$ have the same range as in (a).
In (a) and (c), black color indicates regions where the order parameter vanishes, i.e., synchronization is absent.
When each group synchronizes individually, i.e., in the blue and red regions in (a), the relative phase in (c) takes arbitrary values since there is no fixed phase relation between the two groups.
Parameters:
$V_{AB} = V = 2\gp = \gm$.
}
\end{figure}

The previous analysis was done for a symmetric three-level structure, i.e., $K=0$.
We now vary the asymmetry parameter $K$ in addition to the detuning $\delta$ and present the resulting phase diagram in \cref{fig:2groupsVfix}.
Remarkably, for large $\abs{K}/\gm$, we find synchronization only if $\abs{\delta} \sim \abs{K}$,
while synchronization is absent for $\delta$  around zero.
This is a macroscopic manifestation of the quantum synchronization blockade~\cite{lorchQuantumSynchronizationBlockade2017}:
the two groups synchronize when they are distinct, but not if they are similar.
This is in contrast to the expected behavior that a greater similarity of oscillators increases their tendency to synchronize (see~\cref{fig:2groupsAnh0}(d)).
The absence of synchronization is caused by the discrete energy spectrum of the oscillators.
The effect of the coupling between two oscillators is suppressed when $\abs{K}$ significantly differs from $\abs{\delta}$ because the dominant transitions are off-resonant~\cite{supp}.
Only \textit{close to} the resonances 
$K =\delta$ and $K =-\delta$, there is strong phase alignment of the two oscillators.
This explains the microscopic synchronization blockade, however, it does not yet fully capture the macroscopic quantum synchronization blockade, since we find synchronization only \textit{below} the lines $K = \delta$ and $K = -\delta$.
Below these lines, oscillators of different groups tend to align their phases, while above, they favor opposite phases~\cite{supp}.
This becomes apparent in the phase difference of the two groups, see~\cref{fig:2groupsVfix}(c) which shows the relative phase approaching $\pi$ close to the diagonals.
Each oscillator reacts to the mean fields of both groups, such that their influence cancels if they have opposite phases.
This constitutes another instance of phase frustration which in this case results in
an additional blockade of synchronization for $K>\delta$. 

To summarize the analysis of two groups,
parts of their dynamics can be understood as a typical synchronization transition.
In general, however, the quantum properties change the dynamics significantly:
we demonstrated a blockade of global synchronization
resulting from the quantized nature of the oscillators.
Moreover, an extended blockade of synchronization emerges in the ensemble not present in the case of two coupled oscillators.

\textit{Experimental considerations.---}
Possible experimental realizations include superconducting circuits~\cite{lorchQuantumSynchronizationBlockade2017,Nigg2018} and trapped ions~\cite{lee_QuantumSynchronizationQuantum_2013, Armour2015, lorchQuantumSynchronizationBlockade2017}.
We discuss these two implementations and general requirements to observe the phenomena presented in our work in~\cite{supp}.
We also address finite-size effects, and show that the lifetime of the coherence in a single group increases linearly with the number of oscillators~\cite{supp}.
Since global synchronization can persist for finite-range interactions in networks of classical oscillators~\cite{acebronKuramotoModelSimple2005} and quantum oscillators~\cite{zhuSynchronizationInteractingQuantum2015},
we expect the all-to-all coupling that we assumed not to be essential.

\textit{Conclusion.---}
While quantum effects in synchronization have been studied at the level of few coupled oscillators, it has remained an open question whether these effects remain when increasing the number of oscillators.
To address this issue, we investigated the synchronization behavior of two macroscopically large groups of reactively coupled quantum limit-cycle oscillators.
We demonstrated that quantum effects in synchronization persist on a macroscopic scale:
for a single group, destructive interference manifests itself as a blockade of collective synchronization
if gain and loss rates are comparable;
for two detuned groups of oscillators with an asymmetric level structure, their quantized nature counterintuitively leads to synchronization of dissimilar groups.
We also identified certain aspects of the dynamics that can be understood from classical generic synchronization models:
for a single group, the transition to synchronization necessitates a critical coupling strength to overcome disorder through quantum fluctuations;
for two groups of oscillators with a symmetric level structure in the regime of small inter-group coupling strength, their dynamics can be understood as the synchronization of two classical phase oscillators.
Finally, we uncovered emergent behavior only present in the macroscopic ensemble: phase frustration, i.e., oscillators antialigning their phases, suppresses the global coherence and results in the absence of collective synchronization.

We expect our work to stimulate the exploration of other intriguing aspects in many-body quantum synchronization beyond classical expectations.
Additionally, our results further connect synchronization to 
dissipative quantum phase transitions,
such as superradiance and time crystals.

\vspace{1em}
\begin{acknowledgments}
\textbf{Acknowledgements}
We would like to thank M. Koppenh\"ofer and N. L\"orch for stimulating discussions. Furthermore, we acknowledge the use of QuTiP~\cite{Johansson_2013} and QuantumCumulants.jl~\cite{Plankensteiner2022quantumcumulantsjl}, as well as
financial support from the Swiss National Science Foundation individual grant (grant no. 200020 200481).
\end{acknowledgments}

\pagebreak
\clearpage
\onecolumngrid
\begin{center}
\textbf{\large Supplemental material: Macroscopic quantum synchronization effects}
\end{center}
\newcounter{sfigure}
\renewcommand{\thefigure}{S\arabic{sfigure}}
\stepcounter{sfigure}

\setcounter{table}{0}
\setcounter{page}{1}
\makeatletter
\renewcommand{\theequation}{S\arabic{equation}}

\appendix
\section{1.~Microscopic description}
We present results regarding the microscopic synchronization behavior of three-level quantum oscillators.
These results are used to explain the synchronization behavior observed on a macroscopic scale,
in particular the macroscopic interference blockade and the influence of an asymmetry in the level structure of the oscillators shown in \cref{fig:1group}(c), as well as the quantum synchronization blockade displayed in \cref{fig:2groupsVfix}.

\subsection{1.1.~One oscillator coupled to external drive}
We discuss the microscopic scenario of one three-level system oscillator coupled to an external harmonic drive.
This allows us to explain the macroscopic effects discussed in the main text regarding synchronization of one group (see \cref{fig:1group}),
viz., the existence of an interference blockade for $K\geq 0$ and its absence for $K<0$.

The master equation for the oscillator coupled to an external drive with strength $\Omega$ resonant with the natural frequency of the oscillator is
\begin{align}
\begin{split}
\dot \rho =
    &-i \left[ K \dyad{2} + \Omega (S^+ + S^-),\rho
    \right] +
    \gp \mathcal{D} \left[\dyad{1}{0} \right]\rho
    +
    \gm \mathcal{D} \left[\dyad{1}{2} \right]\rho \, .
    \label{eq:master1}
\end{split}
\end{align}
The external drive has a similar effect as the mean field to which each oscillator is coupled.
In general, it induces a phase preference as indicated by the phase distribution~\cite{lee_QuantumSynchronizationQuantum_2013,rouletSynchronizingSmallestPossible2018}
\begin{equation}
    s(\phi) = \int_0^\pi \mathrm{d} \theta \sin \theta Q(\theta,\phi) - \frac{1}{2\pi} \, .
\end{equation}
Here, we use the Husimi-Q function
\begin{equation}
    Q(\theta,\phi) = \frac{3}{4\pi} \expval{\rho}{\theta,\phi} \, 
\end{equation}
and spin--coherent states
\begin{equation}
    \ket{\theta,\phi} =
    \exp(-i\phi S^z)
    \exp(-i\theta S^y) \ket{2} \, ,
\end{equation}
with the spin--1 operator $S^y = i(S^- - S^+)/2$.
Note that this phase distribution was generalized to $SU(3)$ coherent states in Refs.~\cite{jaseemQuantumSynchronisationNanoscale2020,solanki2022symmetries}, including two free phases. For our discussion, however, it is enough to consider one phase.

The drive represents the influence of the mean field on the oscillator. We choose its strength $\Omega/\gm = 1/10$.
We compute the steady state of the master equation \cref{eq:master1} and show the resulting phase distributions $s(\phi)$ in \cref{fig:1group}(d). 
As discussed in the main text, for negative $K$, the oscillator tends to align in phase with the drive, whereas for positive $K$ it tends to anti-align, hindering synchronization.
For $K = 0$, and $\gp = \gm$, the coherences resulting from the external drive have nearly opposite signs, such that their contributions to the phase distribution partially cancel.
In particular, the first-order contribution $\propto \Omega/\gm$ exactly cancels, and only higher-order contributions remain~\cite{koppenhoferOptimalSynchronizationDeep2019}.
This is the destructive interference blockade: there are two peaks of equal height at $\phi=0$ and $\phi=\pi$ in the phase distribution instead of a single more pronounced maximum that would be expected if the first-order contribution was present.
The first-order contribution is visible in the second panel of \cref{fig:1group}(d), where $\gp = \gm/2$.

\subsection{1.2.~Two coupled oscillators}
\label{app:micro2}
\begin{figure}
    \centering
    \includegraphics[width =6.8in]{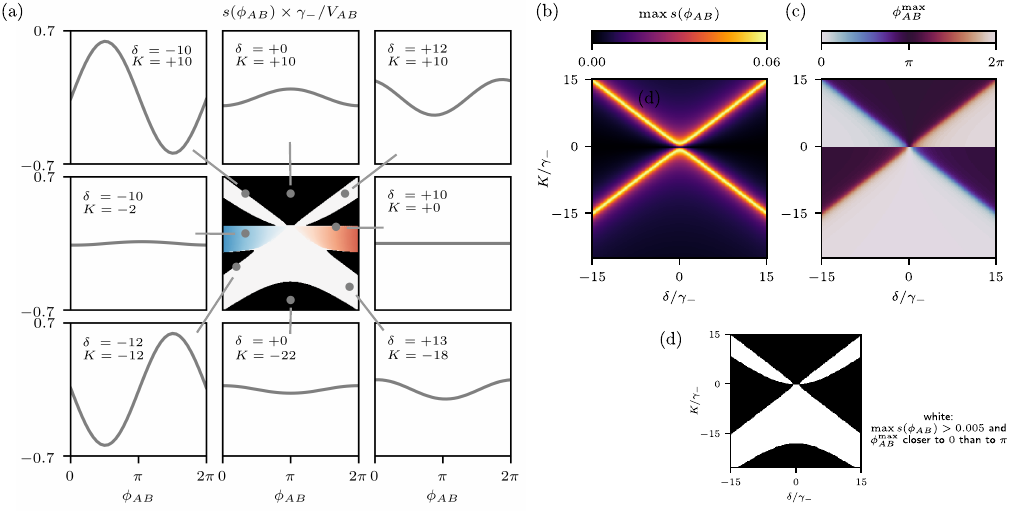}
    \caption{%
    (a) 
        The central panel is the same as \cref{fig:2groupsVfix}(a).
        Each of the eight plots surrounding the central panel shows the distribution of the relative phase, for different values of detuning $\delta$ and level asymmetry $K$ (in units of $\gm$), indicated by a grey dot in the central panel.
    (b) Maximum of the phase distribution. (c) Argument $\phi_{AB}^\mathrm{max}$ of the maximum shown in (b). 
    (d) 
    In this bitmap, white pixels indicate that the maximum of the phase distribution is larger than the threshold of $5\times 10^{-3}$, and the maximum phase is closer to $0$ than to $\pi$.
    The resulting shape closely resembles that of \cref{fig:2groupsVfix}(a).
    Parameters: $\gamma_- = 2\gamma_+$ as in \cref{fig:2groupsVfix}.}
    \label{fig:2group_micro}
\end{figure}

The full quantum master equation (see \cref{eq:system}) for a group size of $N=1$
\begin{align}
\begin{split}
\dot \rho =
    &-i \Bigl[ \frac{\delta}{2}(S^z_A - S^z_B) + K (\dyad{2}_A+\dyad{2}_B) +
    V_{AB} (S^+_A S^-_B + S^+_B S^-_A),\rho
    \Bigr] +
    \\
    &+ \bigl(
    \gp \mathcal{D} \left[\dyad{1}{0}_A \right] + 
    \gm \mathcal{D} \left[\dyad{1}{2}_A \right] +
   \gp \mathcal{D} \left[\dyad{1}{0}_B \right] + 
    \gm \mathcal{D} \left[\dyad{1}{2}_B \right] \bigr) \rho \, ,
    \label{eq:supp_master2}
\end{split}
\end{align}
describes the dynamics of two coupled, detuned oscillators. We omitted the subscript $i=j=1$.
We analyse their steady state using the phase distribution for the relative phase $\phi_{AB}$
\begin{align}
\begin{split}
    s(\phi_{AB}) =
    \int &
    \mathrm{d} \theta_A \mathrm{d} \theta_B  \mathrm{d} \phi_A \mathrm{d} \phi_B \sin \theta_A \sin \theta_B
    \times
    Q(\theta_A,\theta_B,\phi_A,\phi_B) \times \delta(\phi_{AB}-\phi_A+\phi_B) - \frac{1}{2\pi}
    \, ,
    \end{split}
\end{align}
generalizing the Husimi-Q function to two oscillators
\begin{equation}
    Q(\theta_A,\theta_B,\phi_A,\phi_B) = \frac{9}{16\pi^2} \expval{\rho}{\theta_A,\phi_A,\theta_B,\phi_B} \, 
\end{equation}
using the tensor product of spin-coherent states
\begin{equation}
    \ket{\theta_A,\phi_A,\theta_B,\phi_B} =  \ket{\theta_A,\phi_A} \otimes \ket{\theta_B,\phi_B}\, .
\end{equation}
Since the coupling between any two oscillators is small, we use the first-order result for the steady state of the master equation~\eqref{eq:supp_master2} to compute the Husimi-Q function and $s(\phi_{AB})$.
We show the phase distribution for different values of $\delta$ and $K$ in \cref{fig:2group_micro}(a).
\stepcounter{sfigure}
If $\delta = -K$, the energy difference between states $\ket{1}$ and $\ket{2}$ of oscillator $A$ is equal to that between states $\ket{0}$ and $\ket{1}$ of oscillator $B$.
Thus, the transition $\ket{1}_A\otimes \ket{1}_B \leftrightarrow \ket{2}_A \otimes \ket{0}_B$ is resonant which leads to a strongly varying phase distribution.
As an example, we show $s(\phi_{AB})$ for $\delta= -K = -10\gm$ in the top left of \cref{fig:2group_micro}(a).
We also find a strongly varying phase distribution when the energy difference between states $\ket{0}$ and $\ket{1}$ of oscillator $A$ is equal to that between states $\ket{1}$ and $\ket{2}$ of oscillator $B$ (see bottom left plot of \cref{fig:2group_micro}(a) where $\delta=K=-12\gm$). In this case, the transition $\ket{1}_A\otimes \ket{1}_B \leftrightarrow \ket{0}_A \otimes \ket{2}_B$ is resonant.
Note that these two transitions are the most important ones, since the limit-cycle (product) state $\ket{1}_A \otimes \ket{1}_B$ is the most populated one.
The large phase alignment between the two oscillators leads to their synchronization.
When $\abs{\delta}$ differs significantly from $\abs{K}$, the influence of the coupling is suppressed since the dominant transition are off-resonant.
This causes a comparably flat phase distribution (see middle panels on each side of \cref{fig:2group_micro}(a)).
\Cref{fig:2group_micro}(b) shows the maximum of the phase distribution and clearly highlights the resonances as well as the suppression of synchronization between two oscillators when $\delta$ is close to zero for large $\abs{K/\gamma_-}$.
This is the microscopic quantum synchronization blockade.

As discussed in the main text, in the macroscopic ensemble we find an extended synchronization blockade:
In the regions just above the two lines $K=\delta$ and $K=-\delta$, synchronization is also absent.
To explain this we need to consider the preferred relative phase $\phi_{AB}^\mathrm{max} = \mathrm{argmax}_{\phi_{AB}} s(\phi_{AB})$, which we plot as a function of detuning and asymmetry parameter in \cref{fig:2group_micro}(c).
The relative phase is zero below the resonances and $\pi$ above them. At the resonances, they cross $\pm \pi/2$.
The phase shift between oscillators $A$ and $B$ below the resonances is positive for $K\lesssim -\delta$, and negative for $K \lesssim \delta$.
This is reflected in the phase difference of the macroscopic ensembles, see \cref{fig:2groupsVfix}.

Since each oscillator is subject to the mean fields of both groups, the resulting effect can be smaller if both groups tend toward opposite phases, i.e., their mean fields tend toward opposite signs.
We find that $\phi_{AB} = \pm \pi/2$, which is exactly at the resonances, is the threshold where synchronization can still occur.
In the macroscopic ensemble, this results in a phase shift of $\pm \pi$.
If the relative phase of two oscillators is even closer to $\pi$, there is no synchronization.

To highlight the relation to the macroscopic phase diagram, we show in \cref{fig:2group_micro}(d) a bitmap with white indicating both the amplitude surpassing a certain threshold and the phase being closer to $0$ (or $2\pi$) than to $\pi$.
As explained before, these two are reasonable assumptions for the requirements of synchronization of the two groups.
Indeed we find that the resulting white area agrees with the region where the macroscopic system fully synchronizes. 
The value of the amplitude threshold is a free parameter that we chose to be $5\times 10^{-3}$.
Qualitative features, such as the general X-shape whose bottom diagonals are broader than the top ones are independent of this choice.

\section{2.~Supplementary Figures}
\begin{figure}
    \centering
    \includegraphics[width = 6.8in,height = 2.26666666in]{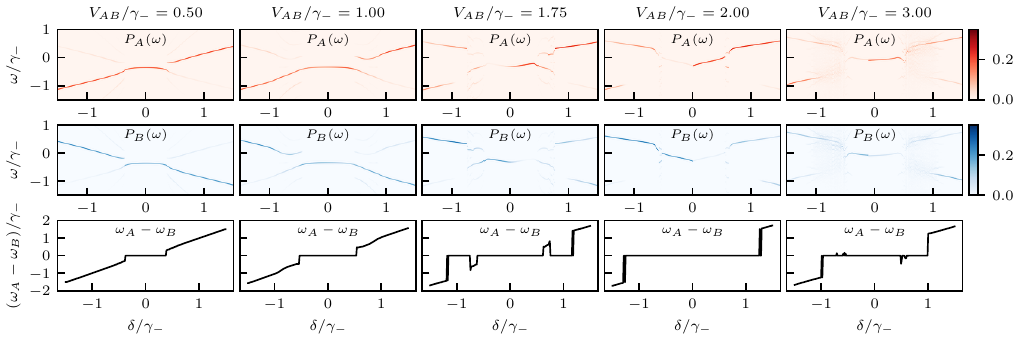}
    \caption{%
    Additional spectra complementing \cref{fig:2groupsAnh0}.
    In each column, for a different value of the inter-group coupling strength $V_{AB}$, the top two panels show the spectra $P_{\sigma}(\omega)$ for $\sigma=A,B$ obtained via Fourier transform of the time evolution of $\expval{S^+}_\sigma$ as a function of detuning $\delta$.
    The bottom panel shows the difference of the two dominant frequencies, $\omega_A - \omega_B$, where $\omega_{\sigma} = \mathrm{argmax}_\omega P_\sigma(\omega)$.
    Parameters as in \cref{fig:2groupsAnh0}: $K=0$, $V=2\gp=\gm$ (such that $V>V_c$).
\label{fig:supp_adler_spectra}}
\end{figure}

\begin{figure}
    \centering
    \includegraphics[width = 6.8in,height = 2.26666666in]{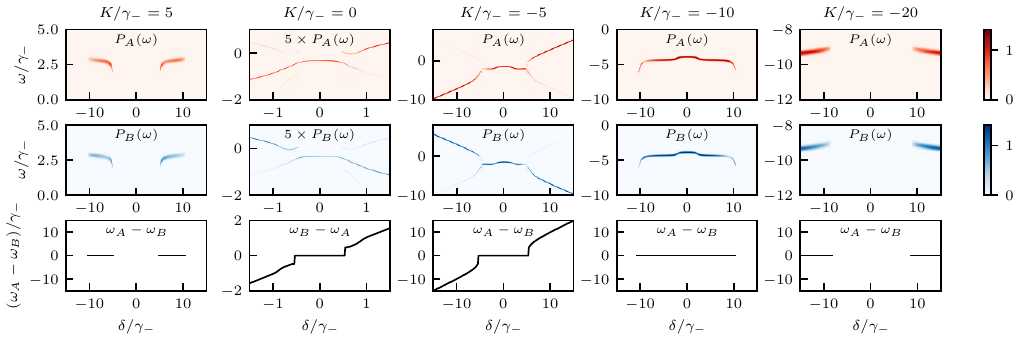}
    \caption{%
    Additional spectra complementing \cref{fig:2groupsVfix}.
    In each column, for a different value of the asymmetry parameter $K$, the top two panels show the spectra $P_{\sigma}(\omega)$ for $\sigma=A,B$ obtained via Fourier transform of the time evolution of $\expval{S^+}_\sigma$ as a function of detuning $\delta$.
    The bottom panel shows the difference of the two dominant frequencies, $\omega_A - \omega_B$, where $\omega_{\sigma} = \mathrm{argmax}_\omega P_\sigma(\omega)$.
    The frequency difference in the bottom row is shown only in those regions where there is a significant peak in the spectra.
    Parameters as in \cref{fig:2groupsVfix}: $V_{AB} = V = 2\gp = \gm$.
    \label{fig:supp_blockade_spectra}}
\end{figure}

In Figs.~\ref{fig:supp_adler_spectra} and \ref{fig:supp_blockade_spectra} we show additional spectra complementing \cref{fig:2groupsAnh0,fig:2groupsVfix}.
In short, they demonstrate that for the results presented in this work, it is not necessary to consider the full spectra $P_\sigma(\omega)$ to characterize the synchronization between two groups, but instead it is  sufficient to use the difference of the dominant frequencies $\omega_{\sigma} = \mathrm{argmax}_\omega P_\sigma(\omega)$.

The full spectra displayed in Figs.~\ref{fig:supp_adler_spectra} and \ref{fig:supp_blockade_spectra} in general show more than one frequency component.
For small inter-group coupling (see Fig.~S2, two left columns, and Fig.~S3, second and third columns), we observe a small additional frequency component that is picked up in each group from the other group.
This additional peak, however, is clearly smaller than the dominant one, and therefore we can safely focus on the main peak.
The possibility of vanishing spectra (see first, fourth and fith column in Fig.~\ref{fig:supp_blockade_spectra}) is also captured by our analysis in \cref{fig:2groupsVfix}, since in this case, the order parameter $\abs{\expval{S^+}_\sigma}$ vanishes as well.
In the regime of the inter-group coupling dominating the intra-group coupling, $V_{AB} > V$ (see Fig.~S2, middle to right columns), however, it is possible to find more than one dominant frequency component.
In this regime, one can also observe the case that one group synchronizes more strongly than the other group.
This is best visible in the fourth column of Fig.~\ref{fig:supp_adler_spectra}, where for negative (positive) detuning $\delta$ group B (A) is more strongly synchronized as indicated by the peak in the spectrum $P_B$ ($P_A$) being more pronounced.
This regime ($V_{AB} > V$) is not relevant for the results presented in our work, and its detailed analysis is left for future study.
\stepcounter{sfigure} 
\stepcounter{sfigure} 

\section{3.~Finite-size analysis}
\begin{figure}
    \centering
    \includegraphics[width = 6in,height = 2.5in]{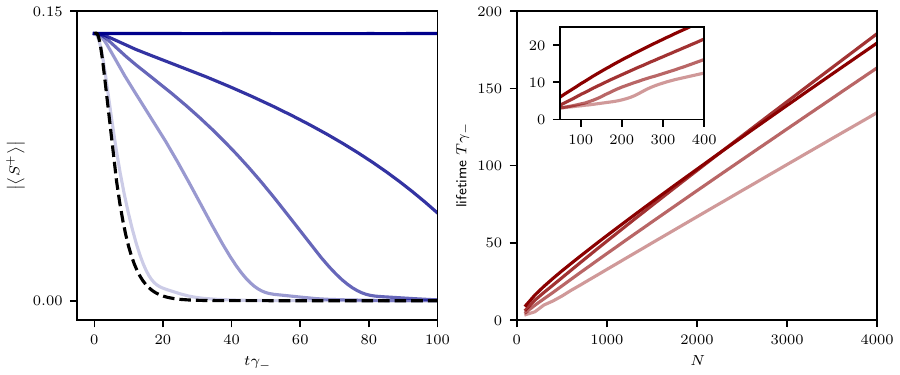}
    \caption{Finite-size analysis.
    Left: The time evolution of the absolute value of the amplitude $\abs{\expval{S^+}}$ for different values of the number of oscillators $N$. From light to darker blue, $N$ takes values $100,500,1000,2000$, and infinity, which corresponds to the mean-field result.
    The black dashed line shows the evolution in the absence of any coupling.
    Right: The lifetime $T\gamma_-$ of the coherence as a function of $N$, for different values of the coupling strength, $V$.
    From dark to light red, $V/\gm$ takes values $0.75,1,1.25,1.5$.
    The inset shows the same data for smaller values of $N$.}
    \label{fig:supp_finite_N}
\end{figure}

To understand the influence of the group size, we go beyond the mean-field treatment and include some correlations between the observables.
A systematic approach is to truncate higher-order cumulants~\cite{cumulants_kubo}.
For this analysis, we truncate at the second-order correlations, i.e., neglect correlations between three and more observables.
To do so, we use the Julia package QuantumCumulants.jl~\cite{Plankensteiner2022quantumcumulantsjl} which provides an automatized way of deriving equations of motion including correlations up to a set order and converting them to Julia functions that can be integrated numerically.
Implementing this for our system of one group, i.e., \cref{eq:system} with $V_{AB}=0$, and time-integrating, we obtain the results shown in \cref{fig:supp_finite_N}.
\stepcounter{sfigure}%
Instead of a persistent non-vanishing value of $\abs{\expval{S^+}}$ that is observed in the mean-field limit $N\rightarrow \infty$, the amplitude decays over time for finite $N$.
We find that the lifetime as measured by the time for the absolute value of the amplitude to decay to $1/e$ increases linearly with the number of oscillators.
The lifetime of the coherence in a group of 500 oscillators reaches $T\gm \approx 30$, four times larger than the lifetime in the absence of coupling.
Due to the linear scaling, reasonable numbers of oscillators are sufficient to observe synchronization in the large network as indicated by the long-lived amplitude $\expval{S^+}$.

\section{4.~Experimental requirements and realizations}

The requirements for experimental observations of the macroscopic quantum synchronization effects discussed in the main text are the following.
All effects predicted in this paper require three-level oscillators subject to gain and loss that generate the limit cycle, and the ability to coherently couple them.
The macroscopic interference blockade can be observed with a single group of identical oscillators, each possessing a symmetric level structure.
On the other hand, the quantum synchronization blockade of two groups 
necessitates an asymmetric level structure and control of the detuning between the two groups.
As discussed in Section 3 of the Supplemental Material, the lifetime of the order parameter that indicates synchronization increases linearly with the number of oscillators.
Therefore, reasonable numbers of hundreds of oscillators are sufficient to observe collective synchronization in the form of long-lived coherences.

An experimental observation of macroscopic quantum synchronization effects is challenging; even synchronization of a few interacting quantum oscillators remains to be experimentally demonstrated.
Here, we propose two possibilities to practically implement large networks of quantum oscillators and hope to stimulate future research towards their experimental realization.
We briefly describe the implementation of the three levels of each oscillator, the gain and loss processes, and the coupling among the oscillators, for two different versatile experimental platforms: superconducting circuits and trapped ions.

In superconducting circuits, following Refs.~\cite{lorchQuantumSynchronizationBlockade2017,Nigg2018}, each oscillator is implemented in a transmon qubit.
Gain and loss are engineered through coupling to ancillary modes, see~\cite{Nigg2018} for details.
These incoherent processes result in a population narrowly distributed around one eigenstate $\ket{n}$, $n>0$, which allows for an effective approximate description of the dynamics using only three levels $\ket{n-1}$, $\ket{n}$, and $\ket{n+1}$.
These correspond to the levels $\ket{0}$, $\ket{1}$, and $\ket{2}$ with gain and loss stabilizing state $\ket{1}$, as considered in the main text.
Since each transmon qubit features an \textit{anharmonic} spectrum, the three levels form an \textit{asymmetric} three-level oscillator that is our building block.
Control of the frequency of the transmon qubits allows for introducing two detuned groups of oscillators.
For the coupling among the oscillators, in principle, each pair of oscillators can be capacitively coupled.
Due to hardware constraints, however, it is difficult to couple all pairs of oscillators in this way.
For collective synchronization, long-range coupling among the oscillators is required.
Such non-local couplings of transmons qubits have been described in~\cite{Majer_2007,Onodera_2020}.

In trapped-ion setups, following Refs.~\cite{lee_QuantumSynchronizationQuantum_2013, Armour2015, lorchQuantumSynchronizationBlockade2017}, each oscillator is implemented in a motional mode of an ion trapped in an anharmonic potential.
Gain and loss can be engineered by employing blue and red sideband transitions that incoherently drive transitions from one to another motional state~\cite{Leibfried_2003}.
The anharmonicity of the energy spectrum allows to address individual transitions with varying strengths by tuning the sideband frequency.
As described above in the case of transmon qubits, by engineering gain and loss to stabilize one motional eigenstate, the dynamics are reduced to this state and the two neighbouring states.
The anharmonic potential results in an asymmetry of the level structure of the three relevant states.
By changing the harmonic part of the potential, the frequency of each oscillator, and hence the detuning between the two groups can be controlled.
All oscillators are naturally coupled via the long-range Coulomb interaction.
The coupling strength can be adjusted by varying the distance of the individual ions.

\section{5.~Supplementary information on methods}
We present additional information on the numerical calculations used to obtain the results of this work ($\mathbb{I}$ denotes the unit matrix). Note that these details, including initial states or integration times, do not influence the results. All spectra are obtained using a discrete Fourier transform after smoothing the amplitudes with a Hann window~\cite{oppenheim1999discrete}.

\Cref{fig:1group}(a): 
The initial state is $\rho_0 = \mathbb{I}/3 + \dyad{1}{2} \times (1+2i)/10 $. The coupling strengths are $V/(\gm+\gp) = 1/5$  and $V/(\gm+\gp) = 3/5$.

\Cref{fig:1group}(b): 
The initial state is $\rho_0 = \mathbb{I}/3 + \dyad{1}{2} \times 1/10 $.
We integrate for a total time of $5000/\gm$, saving the state of the system in $5000$ equally spaced time points.
The time average for the order parameter is performed over the final half of the time points.
The calculation is done for 320 equally spaced values of $V/(\gm+\gp)$.

\Cref{fig:1group}(c):
The initial state is $\rho_0 = \mathbb{I}/3 + \dyad{1}{2} \times 1/10 $.
We integrate for a total time of $10000/\gm$, saving the state of the system in $10000$ equally spaced time points.
The time average for the order parameter is performed over the last $1000$ time points.
The resolution of the grey-scale image is $255\times 255$.

\Cref{fig:2groupsAnh0} and \cref{fig:supp_adler_spectra}:
The initial state for group $A$ is $\rho_0 = \mathbb{I}/3 + \dyad{1}{2} \times 1/10 $, and that of group $B$ is $\rho_0 = \mathbb{I}/3$.
We integrate for a total time of $1000/\gm$, saving the state of the system in $10000$ equally spaced time points.
The last half is used to compute the spectrum.
The resolution of \cref{fig:2groupsAnh0}(c) is $255\times 255$.

\Cref{fig:2groupsVfix}:
The initial state for group $A$ is $\rho_0 = \mathbb{I}/3 + \dyad{1}{2} \times 1/10 $, and that of group $B$ is $\rho_0 = \mathbb{I}/3$.
We integrate for a total time of $500/\gm$, saving the state of the system in $10000$ equally spaced time points.
For the frequency difference (Panel (a)), we use the last half of the samples.
For the order parameter (Panel (b)), we average over the last 1000 samples.
For the relative phase (Panel (c)), we take the values at the final time of integration.
The resolution is $255\times 255$.

\Cref{fig:supp_blockade_spectra}:
Same as for \cref{fig:supp_adler_spectra}, but with total integration time $1000/\gm$.
The second column of \cref{fig:supp_adler_spectra} is the same as the second column of \cref{fig:supp_blockade_spectra}, obtained with total integration time $1000/\gm$.

\end{document}